\documentclass[manuscript]{emulateapj}

\shorttitle{MC translational relaxation of H atoms}
\shortauthors{Panarese \& Longo}

\begin{document}

\title{{Monte Carlo calculation of the translational relaxation
       \\of superthermal H atoms in thermal H$_{2}$ gas}}

\author{A. Panarese %\altaffilmark{5}
}
\affil{IMIP-CNR (section of Bari), via Amendola 122/D, 70126 Bari, Italy\\
       antonio.panarese@ba.imip.cnr.it}
       
\and
       
\author{S. Longo %\altaffilmark{1,2}
}
\affil{Department of Chemistry, University of Bari,via Orabona 4,70126 Bari, Italy\\
       IMIP-CNR (section of Bari), via Amendola 122/D, 70126 Bari, Italy\\
       savino.longo@ba.imip.cnr.it}

\begin{abstract}
A simple and reliable method to study the translational relaxation
of 'hot' H atoms following their production by chemical mechanisms is proposed. 
The problem is relevant to ISM, shocks, photospheres, atmospheric entry problems.
It is shown that the thermalization of H atoms can be conveniently 
studied by a simple Monte Carlo method including the thermal distribution of
background molecules and set the basis for further investigations. 
The transport cross section is determined by the inversion of transport data.
The collisions density of H atoms in H$_{2}$ gas is calculated and discussed also
in the context of simple theories. The application of the results to astrophysical problems is
outlined including numerical results for the reaction H + H$_{2}$O $\rightarrow$ H$_{2}$ + OH. A simple analytical formula for the reaction probability during H atom thermalization is proposed.

\end{abstract}


\section{Introduction}
In many important objects atomic hydrogen H is produced by dissociation processes from
diatomic hydrogen H$_{2}$ or other hydrogen-containing molecules like H$_{2}$O, H$_{2}$S or NH$_{3}$.

Examples are comet comas~\cite{ip1983}, Jupiter's atmosphere~\cite{lodders2011} and photodissociation
regions (PDR)~\cite{hollenbach1999} but the direct dissociation of 
H$_{2}$ by electron impact is important also in shocks as an effect of preheating electrons, and in more general
terms, in all non equilibrium plasmas produced in H$_{2}$ where the electron temperature is much higher than the
gas temperature, a situation than can be produced by electric fields and is easily reproduced in discharge plasmas.

Atoms produced from dissociation reactions are often superthermal, or hot atoms, since 
they gain energy by sliding down a repulsive adiabatic potential to the dissociation limit.
Hot atoms are also produced from symmetric charge exchange reactions, i.e. 
H$^{+}_{fast}$+ H $\rightarrow$ H$_{fast}$ + H$^{+}$ of H atoms with fast ions of different origins.
The average energy of such atoms is a few $eV$. The chemistry of such H atoms is 
therefore a non thermal chemistry, since the usual assumption of a 
Maxwell-Boltzmann distribution of the kinetic energy of H atoms cannot 
be held to relate the average thermal energy and the rate coefficients of collisional processes.

The production and reactions of superthermal H atoms is a 
topic of great astrophysical relevance. Many chemical reactions 
involving H atoms and important for interstellar chemical networks have 
a threshold which can be overcome by the availability of extra kinetic 
energy. Examples are the trapping of H atoms in solid hydrogen~\cite{miyazaky1983},
the hydrogen abstraction from several molecules like H$_{2}$O, H$_{2}$S, hydrocarbons 
and their deuterated versions~\cite{nicholas1989,malcolmelawes1972}, the impact ionization of I group 
metals~\cite{crooks1978}. It is therefore important to develop practical and reliable 
methods to characterize the chemical reactivity of H atoms after their 
production including the effects due to their moderation, thermalization 
and loss in impact and reactions with other species. 

This problem can be tackled by using a continuum slowing down approximation similar to the Fermi
theory~\cite{robson2006}, but a much better method is to apply the rigorous thermalization theory 
developed for the case of neutron kinetics in gaseous moderators~\cite{weinberg1958}.

While the resulting integral equations are very complex, a simpler 
approach to obtain rigorous numerical results is provided by the 
application of a Monte Carlo (MC) method that takes into account the thermal 
distribution of target particles in the collision kernel and the energy 
dependence of the cross section. 

In this paper some calculations of the translational relaxation
and chemical reactivity of superthermal H in H$_{2}$ are presented, with a cross section determined from
the inversion of transport data. An accurate MC model for thermal particle transport 
developed in the past in our group and already validated is applied
to this problem. The appropriate parameter set is individuated.
Results are discussed in the light of several important astrophysical reactions. 

\section{Method of calculation}
Calculations are performed using the MC method for particle transport in a thermal background
described by Longo \& Diomede (2009) and by Panarese et al. (2011).%~\cite{longo2009,panarese2011} 

This method has been recently validated by comparing the calculated values of binary diffusion coefficients
in different gases with calculations based on the Chapman-Enskog development extended to high orders~\cite{panarese2011}.

Although the method is described in the above references, here a self-consistent
short description is provided. The starting point is the expression of the real collision frequency
for a H particle moving with velocity v, given by 
\begin{equation}\label{eqnuno}
 \nu({\bf v}) =\int d^{3} w \; \alpha\left(g\right) f({\bf w}) 
\end{equation}
where $f$ is the velocity distribution function of target particles and $\alpha$ is the collision pair frequency defined as
\begin{equation}
\alpha=\sigma \; g \; n_b.
\end{equation}
$\sigma$ is the total cross section, $g=\left|{\bf{v}}-{\bf w}\right|$ is the relative speed of the collision pair
and $n_b$ is the target particle density.

The method is based on the preliminary selection of a maximum value for the product $g\sigma(g)$ denoted by $(g\sigma(g))_{max}$.
By replacing $g\sigma(g)$ with $(g\sigma(g))_{max}$ in the integral expression (\ref{eqnuno}) this last can be rewritten into the form 
$\nu({\bf v}) = \alpha_{max}$ where
\begin{equation} 
 \alpha_{max}=(g\sigma(g))_{max} \; n_{b}.
\end{equation} 
This replacement implies a potentially non-physical increase of the collision 
frequency, which can be compensated by using the concept of null-collision, 
i.e. the inclusion of artificial scattering events which accounts for the 
difference $\alpha_{max}- \alpha$ but has no effect on the 
motion of H atoms. 

This solution allows an exact simple treatment of collisions in a Test Particle Monte Carlo (TPMC) model. 
This numerical method describes the motion of test particles diluted in a bulk medium of target particles.

In our case, the system is constituted by test particles of H moving in
a H$_{2}$ uniform bulk, in equilibrium at temperature $T_{g}$ and pressure $p$.

Initially test particles are put in the origin of a three-dimensional space and are let to diffuse across the bulk.
Test particles are initialized with the same energy and interact with bulk particles by means of binary collisions.

For each collision, the bulk particle velocity is selected according to the Maxwell-Boltzmann distribution at the temperature $T_{g}$,
using a direct method of sampling. 
For this purpose, setting $v_{i}=r \sin \vartheta$ as the velocity component along the i-direction, a pair of values of $r$ and $\vartheta$ is sampled from $\vartheta=2\pi \eta_{1}$ and $r=(-2kT_{g}\ln \eta_{2}/m_{bulk})^{1/2}$, using two random numbers $\eta_{1}$ and $\eta_{2}$
uniformly distributed between $0$ and $1$.
Finally the value of the i-component of the thermal velocity in the equilibrium bulk is sampled as $v_{i}=r \cos \vartheta$.

In order to remove the extra collision events used to equalize the collision frequency to 
$\alpha_{max}$, a further random number $\eta_{3}$ is compared to the fraction of real collisions given by $\alpha/\alpha_{max}$.
If $\eta_{3}$ is smaller than this quantity, the collision is effective.

After an effective collision, the relative velocity vector must be rotated according to two polar angles, namely $\vartheta$,
the scattering angle, and $\varphi$, the azimuthal angle. This last is uniformly sampled in the interval [0,2$\pi$],
while the selection of $\vartheta$ depends on the interaction model.

Once the scattering angle is known, the scattering is treated taking into account the correlation with the old particle velocity using Euler angles: the relative velocity vector after the collision, $g^{*}$, is calculated as
$$
  g^{*}_{x}=g_{x}\cos\vartheta+B\sin\varphi\sin\vartheta,
$$
\begin{equation}
  g^{*}_{y}=g_{y}\cos\vartheta-B^{-1}\sin\vartheta\left(gg_{z}\cos\varphi+g_{x}g_{y}\sin\varphi\right),
\end{equation}
$$
  g^{*}_{z}=g_{z}\cos\vartheta+B^{-1}\sin\vartheta\left(gg_{y}\cos\varphi-g_{x}g_{z}\sin\varphi\right),
$$
where $B=\left(g^{2}_{y}+g^{2}_{z}\right)^{1/2}$ and $\textbf{g}=\left(g_{x},g_{y},g_{z}\right)$ as above
is the relative velocity before the collision.

$\vartheta$ is determined from a quadrature of the interaction potential $\phi(r)$ based on the known value of the impact parameter $b$
$$
\vartheta(b,E)=\pi-2b \int \limits_{r_m}^{\infty} \left [1 - \frac{b^2}{r^2} - \frac{\phi(r)}{E}\right]^{-1/2}
\frac{dr}{r^2}=\pi - \chi(b,E),
$$
where $r_{m}$ is the distance of closest approach.
The value of  $b$ is obtained from $b = b_{max} \sqrt{\eta_{3}}$. 
$\phi$ is simply given by  $\phi = 2\pi \eta_{4}$.
In case of isotropic elastic scattering, $\cos(\vartheta) = 1 - 2\eta_{5}$.

The motion of the colliding particle of mass $m_{c}$ relative to the bulk target particle of mass $m_{t}$ is equivalent 
to the motion of a particle of mass $\mu=m_{c}m_{t}/\left(m_{c}+m_{t}\right)$ relative to a centre of force. 
The collision energy is calculated as a function of the relative speed $g$ of the interacting pair by $E_{coll}=\mu g^{2}/2$.

As in a binary interaction the centre of mass velocity is a constant, the velocity of the colliding particle
after the collision is given by
\begin{equation}\label{newvelocity}
 {\bf v}^{*}=\frac{m_{c}{\bf v}+m_{t}{\bf w}}{m_{c}+m_{t}}+\frac{m_{t}}{m_{c}+m_{t}}{\bf g}^{*}.
\end{equation}

The time difference between one collision (including null collisions) and the next one is given by the formula 
\begin{equation}
  \Delta t = -\ln \eta_{6}/\alpha_{max}
\end{equation}
where $\eta_{6}$ is again a random number from a uniform distribution, $0<\eta_{6}<1$. This 
time is inversely proportional to the gas density. An appropriate 
parameter to measure the degree of thermalization of a H atom in the 
bulk medium is therefore provided by the product $n\tau$, where $\tau$ is the 
average lifetime of a H atom before its chemical or diffusion loss. 
Calculations performed using different values of $n$ and $\tau$, but leaving 
the value of the product $n\tau$ unchanged, produce the same results. 

The finite lifetime $\tau$ is obtained in the simulation by removing the injected particles with a probability
$1-exp(-\Delta t/\tau)$ before each collision. The removed particle is labelled in such a way that the computer 
simulation is not accounting for it anymore. The simulation proceeds until all particles are removed. 

This procedure is exact when $\tau$ is not dependent on the atom speed and can be used when this approximation is considered feasible.
In specific cases $\tau$ will be determined by the chemical network assumed. In molecular clouds, collisional loss reactions with H$^{+}_{2}$ 
and H$^{-}$ and other species can be important, while in the laboratory H atoms are often lost in reactions with purposely added scavengers
uncommon in space, e.g. iodine atoms. For these collisional losses $\tau = (k_{L} n_{L})^{-1}$ where $k_{L}$ is the appropriate rate coefficient
while n$_{L}$ is the number density of the corresponding reaction partner. The photoionization or the radiation ionization of H atoms are
well described by a constant $\tau$ as well. Heterogeneous recombination on reactor walls in the laboratory and on dust grains in space can
be treated similarly, but the formula must account for diffusion times. In case $\tau$ has a strong dependence on the H atom speed or if a loss
cross section $\sigma_{L}$ is used instead of a loss rate coefficient, the removal process can be improved. For any collision a candidate
collision partner for any loss channel is sampled from a Maxwell distribution. The probability that a collision results in H atom loss
is $p_{L}= \sigma g n_{L} / \alpha_{max}$ and the loss channel is selected if $\eta_{7}<p_{L}$.

Collision events are sampled on a uniform grid on an axis representing the variable $u=\ln(\epsilon/\epsilon_{0})$.
In this way the quantity sampled is the collision density $P(u)du$.
This quantity is normalized to one atom produced per unit time.

\section{Results and applications}
The role of hot atoms in chemical reactions is well known as mentioned in the introduction. With the extra energy provided, these atoms can increase considerably, sometimes by orders of magnitudes, the rate of chemical reactions in which they are involved. There are two ways in which the non equilibrium translational kinetics of H atoms can be described, as a result of calculations by the method described in the previous section. One possibility is to calculate the so-called collision density, or equivalently the energy distribution. The second one is to calculate the rate coefficients of relevant reactions with trace species present in the bulk gas (H$_{2}$).

The collision density $P(\epsilon)$ is defined in such a way that the number of collisions of each atom in the energy range 
$(\epsilon,\epsilon+d\epsilon)$ is given by $P(\epsilon)d\epsilon$. The 
knowledge of this function allows to calculate any collisional rate once 
the probability $w_i(\epsilon)$ of accessing the i-th channel is known. 
Alternatively, the usual kinetic distribution $f(v)$ can be calculated, 
if relevant, by the expression $f(v)\;\nu(v)=P(mv^{2}/2)\;mv$ where $\nu(v)$ 
is the collision frequency for test particles of speed $v$ reported at 
the beginning of the previous section. The collision density has been the 
subject of classical theoretical treatments (see also below)
but in the case of slightly superthermal particles decelerated until 
thermal energy (thermalization), which is relevant here, a full numerical approach is more appropriate.
The collision density has been the subject of stochastic 
calculations, beginning from Rebick \& Dubrin (1970). In their work the authors provide the first 
results for H atoms in Xe, a diluent selected in view of the high molar 
mass and relative chemical inertness. They neglected the effect of the 
target speed in some steps of the calculations, which is a quite 
acceptable approximation for their test case where the mass ratio is 
very high, but not appropriate for H in H$_{2}$. The collision frequency and 
dynamics is exactly accounted using the method of this paper.

\begin{figure}[t]
	\centering
		\includegraphics[scale=0.29]{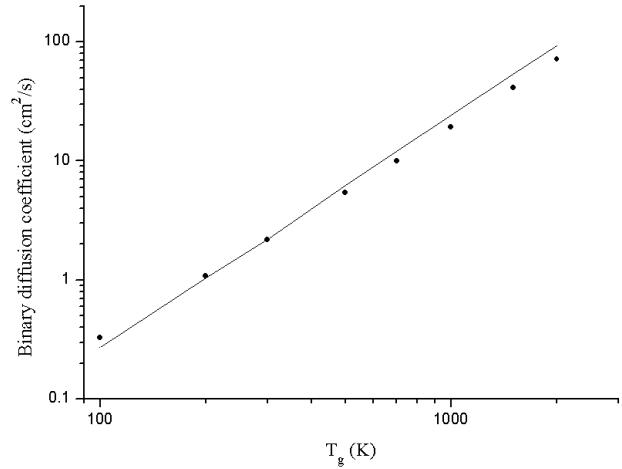}
	\caption{\footnotesize Fit of the binary diffusion coefficient of H in H$_{2}$ calculated
    using data by Stallcop et al. (1996), and using the model cross section in the text.}
	\label{d}
\end{figure}

In order to calculate the frequency of H-H$_{2}$ collisions a reliable momentum transfer cross section is necessary.
Such cross section can be evaluated by fitting the binary diffusion coefficient D of H atoms in H$_{2}$ gas at different temperatures.
This last is calculated using the collision integrals reported by Stallcop et al. (1996) and based on an accurate interaction potential.
The MC method described in the work by Panarese et al. (2011) is used to obtain D as a
function of $T_{g}$ from a guessed cross section of the form $\sigma_{0}(v_{0}/v)^{p}$ 
with $v_{0}=3.75$ $km/s$. The best fit (fig.~\ref{d}) is obtained for $\sigma_{0}=17.5$ \AA$^{2}$ and $p=0.9$.
This values are used in the subsequent calculations of H thermalization.

Fig.~\ref{d} is cut at T$_{g} = 2\times10^{3}$ K, since above this temperature, the equilibrium dissociation of H$_{2}$
at 1atm cannot be neglected. However, the comparison of the two values of the formal quantity D(T$_{g}$) at higher
T$_{g}$ is still meaningful as a check of our simplified cross section at high energy. These two values are actually
very close also at T$_g \simeq 10^{4}$ K.

The numerical parameters revelant to specify the thermalization regime 
in our problems are: the initial H energy $\epsilon_{0}$, the gas temperature $T_{g}$, 
the gas density $n$ and the average lifetime of H atoms in the gas, $\tau$.
This last parameter describes the effect of processes which remove H atoms from the gas. 
This process can be e.g. the photoionization of H atoms. In this and similar cases $\tau$ is defined as
usual through the photochemical expression $\tau = n_H/Rate$ where $Rate$ is the chemical loss rate of H atoms. 

In order to reduce the number of parameters, in this study $\epsilon_{0}$ is fixed to $2.4\;eV$. 
This value is compatible with direct dissociation from the $^{1}\Sigma_{g}$
ground state to the triplet state correlated to H(1s) + H(1s) and provides an important example.
This choice of a delta function for the energy distribution of the source can be extended
to more complex cases by simple statistical sampling of the more involved source function.

\begin{figure}[t]
	\centering
		\includegraphics[scale=0.30]{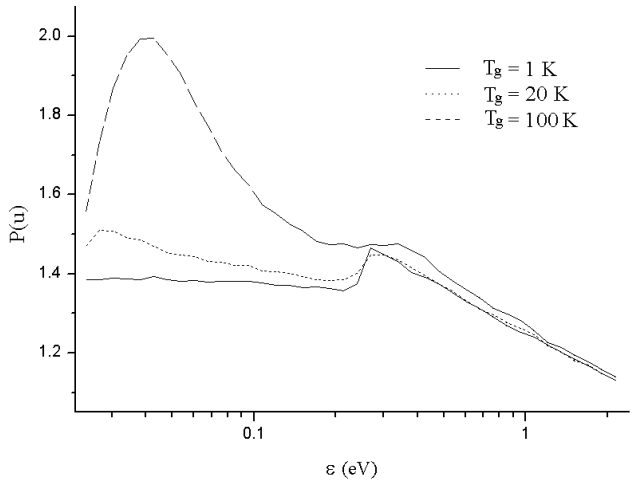}
	\caption{\footnotesize Solution in the cases of T$_{g}$=1K, T$_{g}$=20K and T$_{g}$=100K. The Placzek function, characterized by a discontinuity 
	for $\epsilon$=$\epsilon_{0}$/9, is evident. The singular contribution at $\epsilon=\epsilon_{0}$ (see text) is not represented.}
	\label{unk}
\end{figure}

The collision density in case of H atoms originated with a single energy is characterized by a singular contribution corresponding
to the first collision of H with H$_{2}$ with a H energy by necessity equal to $\epsilon_{0}$.  A finite amplitude for $P(u)$ could
be obtained by selecting the initial energy from a Gaussian like in the work by Prisant et al. (1978).  %~\cite{prisant1978}.
Following the custom for $P(u)$ plots in nuclear applications~\cite{weinberg1958} this contribution have not represented.

Results for $P(u)$ are better discussed in the light of the semi-analytical theory of Placzek (1946),
based on a iterative solution of a simplified integral equation, which is obtained for $P(u)$ in the
limit of zero gas temperature and for rigid sphere elastic scattering.
It is important to note that stochastic approaches allow to remove both limitations of Placzek equation. Nevertheless, the
solution of this equation allows to establish three important general features which are approximately valid also for $T_{g}>0$:

\begin{enumerate}
	\item In the so called asymptotic region far below $\epsilon_{0}$ but close to the bulk thermal 
        energy, $P(u)=1/\xi$ where $\xi$ is the average logarithmic energy loss for single 
        collision. For H in H$_{2}$, this means $P(u)=1.4$. 
	\item Very close to the source energy $\epsilon_{0}$, $P(u)$ is given by a simple expression as a
       	function of the mass ratio $A =m_{H_{2}}/m_{H}=2$ here, i.e. $P(u)=(A+1)^2/4A = 1.125$. 
	\item A discontinuity is present in the function at $\epsilon = \epsilon_{0}((A-1)/(A+1))^2 = 0.11 \epsilon_{0}$
	      associated to the lowest energy value of the initial redistribution by the first collision.
	      In fact, the maximum loss for the H energy corresponds to a head-on collision with an H$_{2}$ at rest:
	      in this case the final H speed is equal to $(A-1)/(A+1)=1/3$ the initial H speed.
\end{enumerate}
Such features are confirmed by our calculations for different values of $T_{g}$, as shown in fig.~\ref{unk}.
In this case the particles were removed when moderated below $0.024 \; eV$ in order to avoid any interference
in the moderation kinetics. This choice produces a non-exponential loss of the particles during moderation.

\begin{figure}[b]
	\centering
		\includegraphics[scale=0.28]{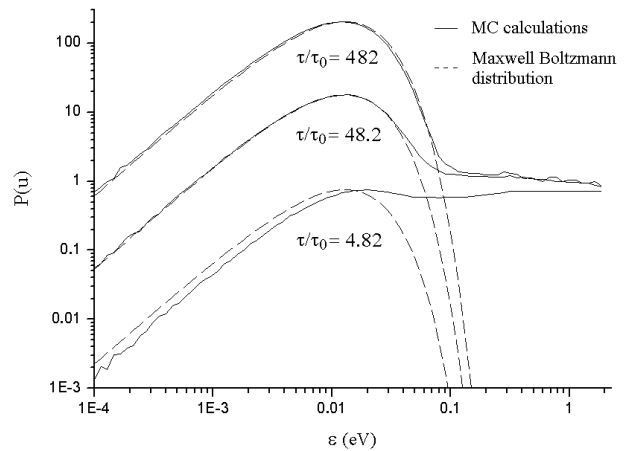}
	\caption{\footnotesize Collision density vs particle energy in the case T$_{g}$=100K for various values of $\tau/\tau_{0}$.
	                       The singular contribution at $\epsilon=\epsilon_{0}$ (see text) is not represented.}
	\label{centok}
\end{figure}

In fig.~\ref{centok} the collision density corresponding to $T_{g}=100K$ and different $\tau$ is reported.
The equilibrium result is also reported for reference. The plots illustrate the role and effects of the two 
fundamental parameters $\tau/\tau_{0}$ and $T_{g}$ on the calculated collision density. 
$\tau_{0}=(n\sigma_{0}v_{0})^{-1}$ is the characteristic collision time.
An estimate of $\tau_{0}$ for a typical ISM density of 10$^{3}$ cm$^{-3}$ leads to $\tau_{0} \simeq 1.5 \times 10^{6} \;$s.

A superthermal tail is associated to the slowing down of 
freshly produced H atoms, whose distribution shape is affected by $T_{g}$ and the 
shape of $\sigma_{g}$ but essentially corresponds to the Placzek solution for $A=2$.
The low energy component of $P(u)$ is essentially determined by the equilibrium contribution, 
the relative importance of the two being controlled by the parameter $\tau/\tau_{0}$.
Even at energies much higher than those corresponding to the 
Maxwellian bulk, in the tail region, the thermal distribution of 
molecules cannot be neglected. This has the effect of smearing the 
Placzek peak of fig.~\ref{unk} which does not appear anymore as a salient 
feature at temperatures higher than a few 10's K even for rigid sphere scattering.
Only detailed calculations can then establish accurately the tail shape.

The presence of the superthermal tail is expected to limit the validity of the traditional chemical 
kinetics of astrophysical objects and phenomena in several ways. For 
example, the rate coefficient of a reactive process involving H atoms 
will not be given by the equilibrium formula based on the Maxwell-Boltzmann distribution, but a superthermal contribution to
the rate coefficients will need to be accounted for, i.e. $K_{tot} = K_{eq}(T_g)+K_{neq}$~\cite{ip1983,crooks1978}.
This last contribution $K_{neq}$ can be determined in specific cases by running a MC simulation based on the 
prescriptions provided in this paper. 

Details differ depending on the gas temperature and the threshold of
the reaction. For reactions with no threshold (where effects are still 
possible due to cross section shape or resonances) or a threshold 
comparable to the thermal energy, the rate coefficient can be corrected by 
the modification of the translational distribution which is not described by a Maxwell-Boltzmann law.
The expression of the rate coefficient K$_{p}$ for a process p, whose related cross section
is $\sigma_{p}$, is a functional of the H translational distribution 
\begin{equation}\label{eqnsette}
 K_{p}=\sqrt{\frac{2}{\mu}}\int^{\infty}_{0}\sigma_{p}(\epsilon)\epsilon^{1/2}f_{cm}(\epsilon)d\epsilon,
\end{equation}
where  f$_{cm}(\epsilon)$ is the distribution of impact energy in the c.m. frame.
In the equilibrium case f$_{cm}(\epsilon)$ is a Maxwellian distribution with m replaced by $\mu$ and K$_{p}$ is a function of T$_{g}$.
When a superthermal component is present, this equation is still valid but K$_{p}$ cannot be written as a function of the gas temperature.
Examples of reactions which may be reconsidered in this light are the radiative association reactions of H with
C$^{+}$~\cite{barinovs2006} and H$^{+}$~\cite{stancil1993} which are of importance for the interstellar medium
and for the early Universe chemistry respectively.

Superthermal atoms can also affect the cooling function of the gas, due to the strong energy dependence
of the collisional deexcitation coefficients, i.e. the rate constants of the reactions 
H+H$_{2}(v,j)\rightarrow$ H+H$_{2}(v',j')$~\cite{capitelli2006}.
The complex thermalization of H atoms will 
also affect preheating in strong shocks in hydrogen. The atomic 
component is more mobile than the bulk gas because of the much higher 
diffusion coefficient, and can contribute to affect the shock profile.
A similar effect has been studied in the past by Bruno et al. for nitrogen 
shocks by a Direct Simulation Monte Carlo (DSMC) model~\cite{bruno2002}.

A much more important and complete nonthermal effect is expected in cases where the reaction threshold 
is much higher than the average thermal energy: the thermal rate of such 
processes is essentially zero, and calculations based on the method 
described in this paper can be used to define a specific rate $Q$ which enters in an expression like
\begin{equation}\label{eqnotto}
 Reaction \; rate = Q \chi_{s} dn_{H}/dt
\end{equation}
where $\chi_{s}$ is the mole fraction of the trace partner $s$ and $dn_{H}/dt$
is the production rate of high energy atoms per unit volume and time. 

$Q$ is therefore the number of reactive collisions of H atoms with a reactive trace species during their moderation,
normalized to unit fraction of the trace (but in practice the fraction is $<<1$ being $s$ a trace).

Examples of such reactions are the reaction of hydrogen with
H$^{-}$ leading to detachment in the so-called non associative channel:
\begin{equation}
  \mbox{H}^{-}+\mbox{H} \longrightarrow 2\mbox{H} + \mbox{e}.
\end{equation}
This process has a threshold of about $0.7 \; eV$ and its cross section increases
by an order of magnitude in the energy range $1-10\;eV$~\cite{reiter2003}.
This reaction can be of relevance for chemical networks connected with radiation transport in
astrophysical systems in view of the great importance of H$^{-}$ ions as radiation absorbers. 

Another example is the abstraction of H atoms from water, e.g. the 
reaction
\begin{equation}
  \mbox{H}+\mbox{H}_{2}\mbox{O} \longrightarrow \mbox{H}_{2}+\mbox{OH}.
\end{equation}

This is potentially an important reaction in astrophysical systems, since it represents a source of hydroxyl radicals in 
rotovibrationally excited states. In spite of this, the above reaction is usually not included in chemical networks in view of
its relative high threshold, about $1\;eV$. This situation, however, can change in presence of superthermal atoms, and for future
reference a calculation of its rate coefficient is provided as a function of the initial energy of H atoms.

The rate constant $Q(\epsilon_{0},T_{g})$ can be evaluated by calculating, in a MC simulation, the sum 
\begin{equation}
 Q(\epsilon_{0},T_{g})=N^{-1} \sum \frac{g_{H/H_{2}O}}{g_{H/H_{2}}} \frac{\sigma_{abs}(g_{H/H_{2}O})}{\sigma_{el}(g_{H/H_{2}})}
\end{equation}
where $\sigma_{abs}$ is the abstraction cross section and $\sigma_{el}$ is the cross section of the elastic
process, N is the number of H atoms in the simulation and the sum includes all H/H$_{2}$ elastic collisions.

The cross section $\sigma_{abs}(\epsilon)$ is that reported as a full line in fig.4 in the paper by Brouard et al. (2004).

\begin{figure}[b]
	\centering
		\includegraphics[scale=0.28]{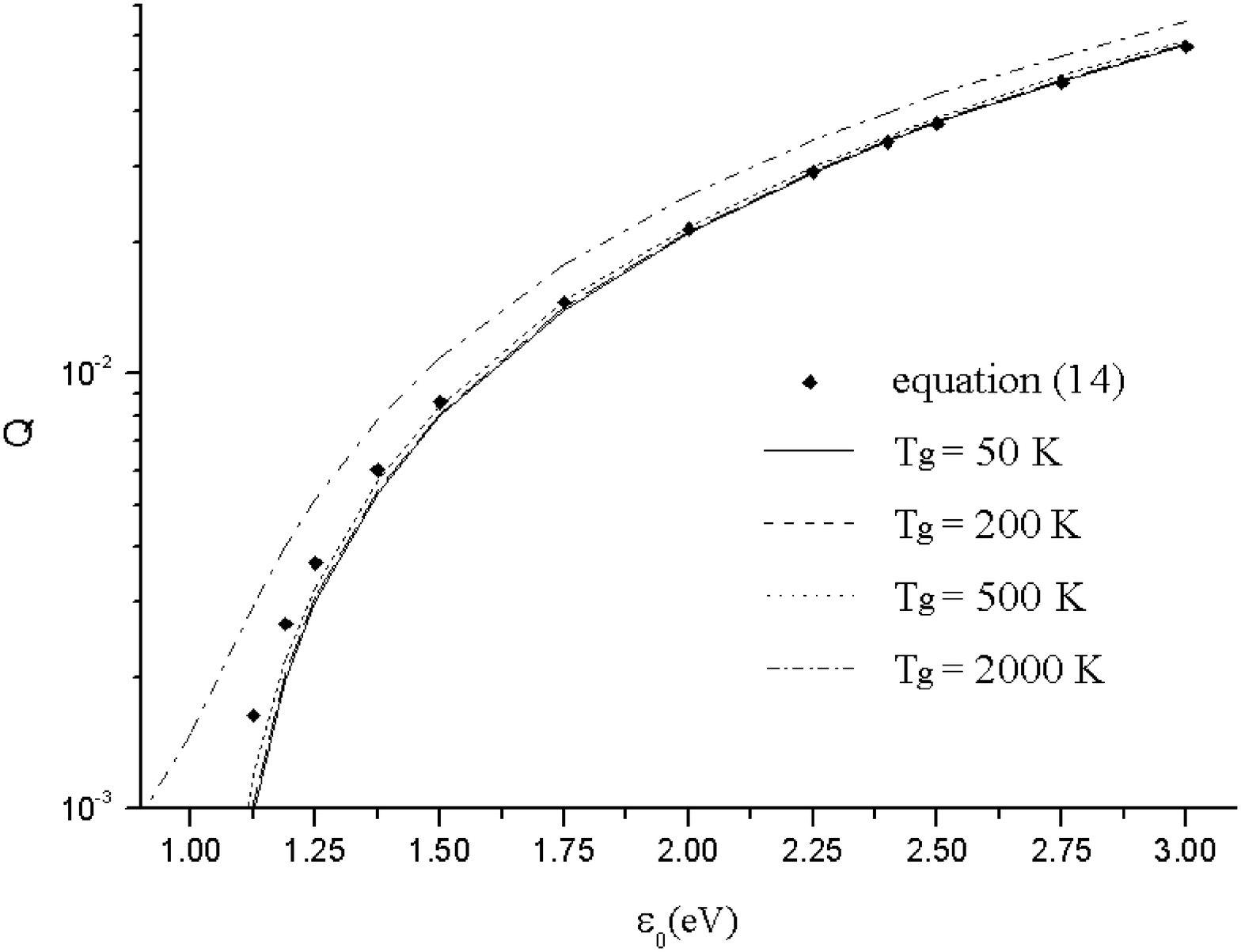}
	\caption{\footnotesize Q vs particle energy in the case $T_{g}=50K$, $T_{g}=200K$, $T_{g}=500K$ and $T_{g}=2000K$.}
	\label{cinquantakdieci}
\end{figure}

The results in fig.~\ref{cinquantakdieci} show that hot H atoms can be an effective source of OH radicals in presence of
traces of water diluted in H$_{2}$. The slight increase observed for the highest value
of T$_{g}$ ($2\times 10^{3}$K) is due to the effect of the relative speed in H/H$_{2}$ collisions on the energy distribution
of H atoms and confirms the necessity of calculating accurately the effects of the relative speed under some circumstances. 

The use of the MC method described here produces an exact evaluation of the quantity Q. An approximate analytical estimate can
be provided under the hypotheses that:
\begin{enumerate}
	\item the Placzek equation applies;
	\item the reactive cross section in the energy range $(0,\epsilon_{0})$ is larger for $\epsilon$ close to $\epsilon_{0}$;
	\item the mass of the H atom is much smaller than that of the reaction partner.
\end{enumerate}
The last hypothesis allows to neglect the difference between the energy of the atom and the energy available
for the collision. An analytical solution of the Placzek equation in the energy range $(\alpha \epsilon _{0},\epsilon _{0})$,
where $\alpha = ((1-A)/(1+A))^2$, is given by the expression~\cite{glasstone1952}:
\begin{equation}
P(\epsilon) = \delta(\epsilon-\epsilon _{0}) + \frac{\epsilon_{0}^{\alpha/(1-\alpha)}}{1-\alpha} \frac{1}{\epsilon^{1/(1-\alpha)}}.
\end{equation}
Having defined the function
\begin{equation}
q(\epsilon) = \sigma(\epsilon)/\sigma_{el}(\epsilon),
\end{equation}
Q can be written as the result of the integration in the range $(0,\epsilon_{0})$ of the product qP, i.e.
\begin{equation}\label{eqnquattordici}
Q = q(\epsilon_{0}) + \frac{\epsilon_{0}^{\alpha/(1-\alpha)}}{1-\alpha} \int^{\epsilon_{0}}_{KT_{g}} q(\epsilon)\frac{d\epsilon}{\epsilon^{1/(1-\alpha)}},
\end{equation}
where the lower integration limit is a reasoble cutoff (see fig.\ref{centok}), since at such low
 energies the collision density is thermal.

In fig.~\ref{cinquantakdieci} results of eq.~(\ref{eqnquattordici}) for several values of $\epsilon_{0}$ are reported for the reaction of
H abstraction from water. In this case the agreement is very good, since this reaction meets well the requirements 2 \& 3 above.
Of course the results at the highest value of T$_{g}$ are not matched since at high T$_{g}$ the Placzek equation is not exact anymore.
Eq.~(\ref{eqnquattordici}) can be used as an alternative to MC calculations for fast, approximate calculations in other cases.

\section{Conclusions}
In this paper it has shown that the problem of the thermalization of 
hot H atoms produced by photochemical, electron or ion impact processes in 
hydrogen based plasmas of astrophysical relevance can be conveniently 
addressed by a simple MC procedure recently developed, which, while being 
very simple, allows for a rigorous treatment of the thermal 
distributions of background species. The relevant physical parameters are identified and
the importance of superthermal H tails for the chemical network and thermal balance
in different systems is discusswd. As an example the abstraction
probability of H atoms with water molecules is evaluated as a function of the initial H
energy $\epsilon_{0}$ and $T_{g}$. A simple formula is proposed to estimate the non thermal contribution
to a reaction rate.

\section*{Acknowledgements}
Work partially supported by MIUR-Universit\'a degli Studi di Bari ("fondi di Ateneo 2011") and EU - FP7 project "Phys4Entry".
The authors are grateful to the anonymous referee for the helpful suggestions.

\clearpage

\end{document}